\documentclass[conference]{IEEEtran}
\IEEEoverridecommandlockouts

\usepackage{cuted}
\usepackage{cite}
\usepackage{amsmath,amssymb,amsfonts}
\usepackage{algorithmic}
\usepackage{graphicx}
\usepackage{textcomp}
\usepackage{xcolor}
\def\BibTeX{{\rm B\kern-.05em{\sc i\kern-.025em b}\kern-.08em
    T\kern-.1667em\lower.7ex\hbox{E}\kern-.125emX}}
\begin{document}

\title{Performance Analysis of MDI-QKD in Thermal-Loss and Phase Noise Channels
}
\author{\IEEEauthorblockN{1\textsuperscript{st} Heyang Peng}
\IEEEauthorblockA{\textit{University of Luxembourg} \\
\textit{The Interdisciplinary Center for}\\
\textit{Security Reliability and Trust (SnT)}\\
Luxembourg \\
https://orcid.org/0009-0005-1754-4500}
\and
\IEEEauthorblockN{2\textsuperscript{nd} Seid Koudia}
\IEEEauthorblockA{\textit{University of Luxembourg} \\
\textit{The Interdisciplinary Center for}\\
\textit{Security Reliability and Trust (SnT)}\\
Luxembourg \\
https://orcid.org/0000-0002-7533-6778}
\and
\IEEEauthorblockN{3\textsuperscript{nd} Leonardo Oleynik}
\IEEEauthorblockA{\textit{University of Luxembourg} \\
\textit{The Interdisciplinary Center for}\\
\textit{Security Reliability and Trust (SnT)}\\
Luxembourg \\
https://orcid.org/0009-0005-6919-9841}
\and
\IEEEauthorblockN{4\textsuperscript{rd} Symeon Chatzinotas, IEEE Fellow}
\IEEEauthorblockA{\textit{University of Luxembourg} \\
\textit{The Interdisciplinary Center for}\\
\textit{ Security Reliability and Trust (SnT)}\\
Luxembourg \\
https://orcid.org/0000-0001-5122-0001}
}
\maketitle

\begin{abstract}
Measurement-device-independent quantum key distribution (MDI-QKD), enhances quantum cryptography by mitigating detector-side vulnerabilities. This study analyzes MDI-QKD performance in thermal-loss and phase noise channels, modeled as depolarizing and dephasing channels to capture thermal and phase noise effects. Based on this channel framework, we derive analytical expressions for Bell state measurement probabilities, quantum bit error rates (QBER), and secret key rates (SKR) of MDI-QKD. Our simulations reveal that SKR decreases exponentially with transmission distance, with performance further degraded by increasing thermal noise and phase noise, particularly under high thermal noise conditions. These findings offer insights into enhancing MDI-QKD's noise resilience, supporting secure key generation in practical, noisy environments.

\end{abstract}


\section{Introduction}
Quantum key distribution (QKD) enables secure key sharing between two parties, Alice and Bob, ensuring encrypted communication with total confidentiality guaranteed by the laws of quantum physics \cite{b1, b2}. The first QKD protocol, BB84, introduced by Bennett and Brassard, utilized discrete variables (DV) with single-photon states, establishing a robust framework for secure communications \cite{b1}. Subsequent developments extended QKD to continuous-variable (CV) protocols, such as the squeezed-state protocol, leveraging entangled states for enhanced performance \cite{b3, b4, b5}.
However, pratical transmission media, such as optical fibers, introduce thermal noise and phase noise, which arise from physical phenomena like random photon scattering due to thermal fluctuations in the fiber material and phase drifts caused by environmental pertubations such as temperature variations or mechanical vibrations. These degrade QKD performance by reducing secret key rates (SKR) and increasing quantum bit error rates (QBER), posing significant challenges to practical deployment \cite{b6, b7, b8}. While studies have explored thermal noise and phase noise effects on both DV- and CV-QKD protocols, research indicates that CV-QKD is robust in low-to-moderate loss regimes, whereas DV-QKD performs better in high-loss scenarios \cite{b9, b10, b11, b12}. The specific impact of these noise sources on Measurement-Device-Independent QKD (MDI-QKD), however, remains underexplored.
MDI-QKD, a DV protocol, eliminates detector-side vulnerabilities, making it a cornerstone for practical quantum networks where a secure long-distance key distribution is critical \cite{b13, b14, b15, b16, b17}. Given its pivotal role in enabling secure quantum communication across metropolitan or satellite-based networks \cite{b18}, understanding MDI-QKD’s behavior under thermal noise and phase noise is essential.

 In this work, we model thermal-loss and phase noise channels as depolarizing and dephasing channels to capture thermal ($N_{\text{th}}$) and phase noise ($\sigma_{\theta}$) effects. The depolarizing channel captures thermal noise's uniform state mixing across all polarization bases, critical for SKR impacts, while the dephasing channel isolates phase noise's coherence loss in the X-basis, vital for eavesropping detection in MDI-QKD, thus, we chose this combined models. Assuming ideal single-photon sources and perfect detectors—eliminating dark counts and other practical imperfections—we derive analytical expressions for projection probabilities onto Bell states, QBER, and SKR, focusing on the channel’s impact on system performance. Through numerical simulations under these idealized conditions, we evaluate MDI-QKD performance across varying noise levels and transmission distances. Our study provides a theoretical framework to optimize MDI-QKD's noise resilience, addressing key challenges for its deployment in noisy quantum communication systems and supporting its application in secure quantum networks.

The paper is structured as follows. In Section~\ref{Sec2}, we establish the theoretical and modeling framework for the thermal-loss and phase noise channels. Section~\ref{sec3} presents the MDI-QKD analysis in terms of SKR and QBER under the influence of thermal-loss and phase noise. Section~\ref{sec4} presents numerical simulations evaluating the performance of MDI-QKD in the presence of these noise effects. Finally, Section~\ref{sec5} concludes the paper.
\section{Thermal Noise and Phase noise in MDI-QKD}
\label{Sec2}
\begin{figure}[t!]
\centerline{\includegraphics[width=0.5\linewidth]{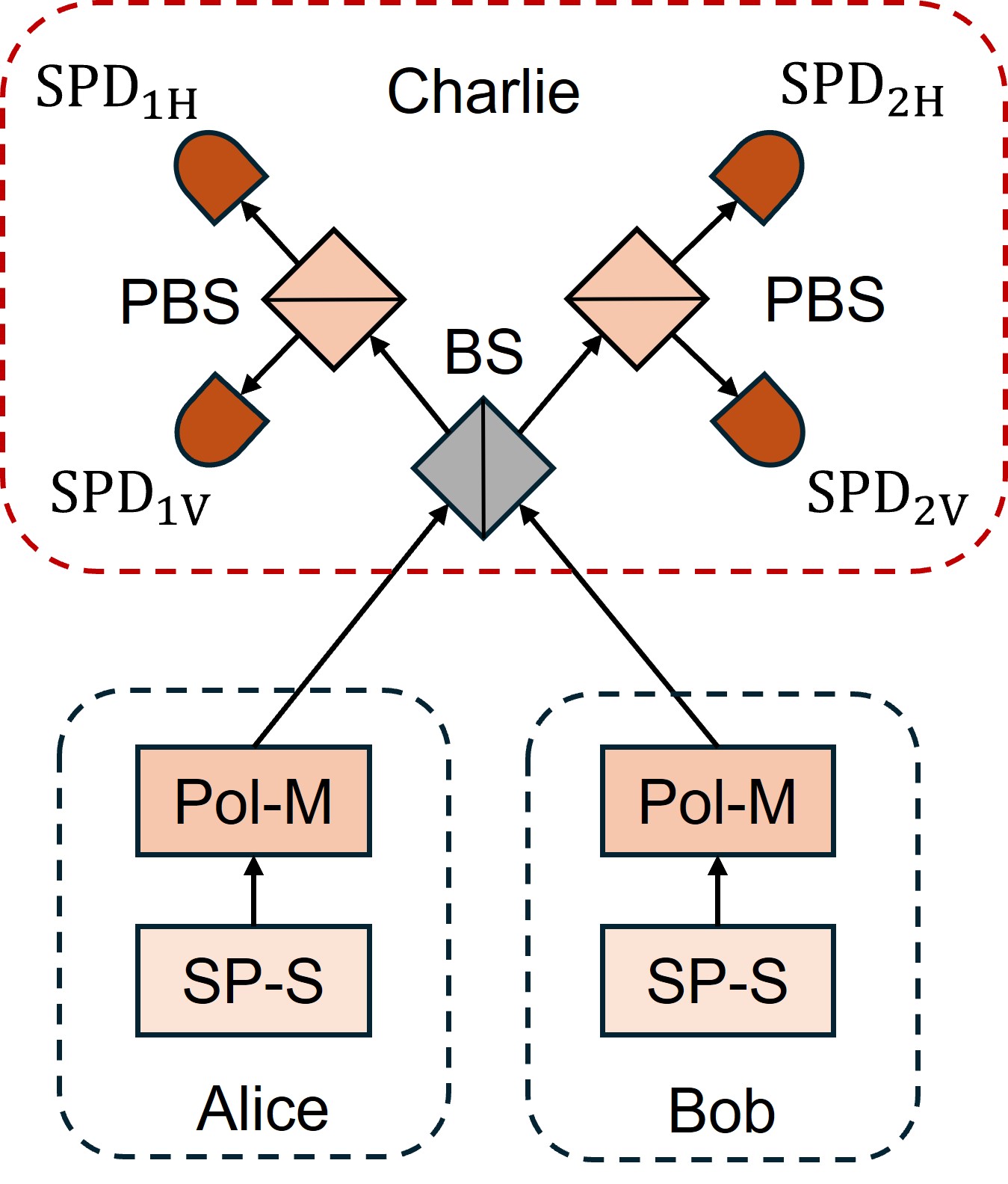}}
\caption{MDI-QKD protocol, SP-S: Single Photon Source, Pol-M: Polarization Modulator, BS: Beamsplitter., PBS: Polarization Beamsplitter, SPD: Single Photon Detector }
\label{fig1}
\end{figure}

In MDI-QKD, Alice and Bob transmit single-photon states through optical fiber channels to an untrusted third party, Charlie, as shown in Figure.~\ref{fig1}. In realistic scenarios, these fiber channels are subject to  thermal-loss and phase noise -- two key factors that can compromise security if not properly adressed. To accurately model these effects, we treat the channels as a combination of depolarizing and dephasing channels. These models capture the essential quantum characteristics of the communication channels and provide a solid theoretical foundation for analyzing the impact of noise. In the following, we present the channel models and their derivations in detail.
\subsection*{1. Thermal-loss Channel}

Thermal-loss channels are common environmental disturbances in QKD, especially in fibre-optic transmission, introduction of background thermal noise denoted by \( N_{\text{th}} \) due to scattering and absorption. To model this effect, we adopt the approach in \cite{b6}, treating the thermal-loss channel as equivalent to a depolarizing channe, as it effectively captures the uniform randomization of photon polarization states across all bases. The depolarizing channel acts on a single-qubit density matrix \(\hat{\rho}\) as follows: 

\begin{equation}
\hat{\rho}' \to (1 - \lambda) \hat{\rho} + \frac{\lambda}{2} \mathbb{I}
\label{eq:depolarizing}
\end{equation}
where \( \mathbb{I} = |H\rangle\langle H| + |V\rangle\langle V| \) is the identity operator in the polarization basis. The depolarization parameter \(\lambda\) quantifies the extent of noise-induced mixing. Its derivation stems from a beamsplitter model, where Alice’s input state is a single photon \( |1\rangle_A \), with the environment in a thermal state:

\begin{equation}
\hat{\rho}_{\text{th}} = \sum_{n=0}^\infty \frac{N_{\text{th}}^n}{(1 + N_{\text{th}})^{n+1}} |n\rangle\langle n|_E
\label{eq:thermal_state}
\end{equation}
with  \(N_\text{th}\) being  the average number of thermal noise photons. For an input state \( |1, n\rangle_{AE} \), the output state \( \hat{\rho}_{AF}' \) is traced over the environment mode \( F \), yielding the unnormalized output \( \hat{\rho}_A' \):

\begin{equation}
\hat{\rho}_A' = \frac{\eta}{\gamma^4} \hat{\rho}_A + \frac{N_{\text{th}}(1 + N_{\text{th}})(1 - \eta)^2}{\gamma^4} \mathbb{I}
\label{eq:rho_A_prime}
\end{equation}
where \(\eta\) is the channel transmissivity and\(\quad \gamma = 1 + N_{\text{th}} - N_{\text{th}} \eta \)  is the normalization factor dependent on \(N_\text{th}\) and \(\eta\).

The single-photon success probability, or the likelihood of detecting a single photon at Charlie, is: 

\begin{equation}
P_S = \text{Tr}(\hat{\rho}_A') = \frac{\eta + 2 N_{\text{th}}(1 + N_{\text{th}})(1 - \eta)^2}{\gamma^4}
\label{eq:success_prob}
\end{equation}
As a consequence, the normalized conditional density matrix is:

\begin{equation}
\hat{\rho}_A' / P_S = (1 - \lambda) \hat{\rho}_A + \frac{\lambda}{2} \mathbb{I}
\label{eq:normalized_rho}
\end{equation}

with:

\begin{equation}
\lambda = \frac{2 N_{\text{th}}(1 + N_{\text{th}})(1 - \eta)^2}{\eta + 2 N_{\text{th}}(1 + N_{\text{th}})(1 - \eta)^2}
\label{eq:lambda}
\end{equation}

This model illustrates how thermal noise \( N_{\text{th}} \) and channel transmissivity \( \eta \) jointly determine the depolarization and success probability of single-photon transmission. Since Alice and Bob operate through independent channels, with parameters \( \lambda_A, P_S^A \) and \( \lambda_B, P_S^B \).

\subsection*{2. Phase Noise Channel}

Phase noise arises from environmental perturbations, such as temperature fluctuations or mechanical vibrations, which disrupt the phase coherence of quantum states. In MDI-QKD, this noise particularly impacts the X-basis, where phase information is critical to detect eavesdropping attempts. We represent phase noise as a dephasing channel as in \cite{b6}, which selectively affects the non-diagonal elements of the density matrix. The phase noise channel transforms the density matrix as:

\begin{equation}
\hat{\rho} \to \begin{bmatrix} \rho_{00} & \bar{r}^2 \rho_{01} \\ \bar{r}^2 \rho_{10} & \rho_{11} \end{bmatrix}
\label{eq:phase_noise}
\end{equation}
where \( \bar{r}^2 = e^{-\sigma_\theta^2} \), and \( \sigma_\theta \) is the standard deviation of the phase noise. This effect originates from a random phase rotation, modeled as: 

\begin{equation}
\hat{\rho} \to \int_{-\pi}^\pi f(\theta) e^{i \hat{n} \theta} \hat{\rho} e^{-i \hat{n} \theta} d\theta
\label{eq:phase_rotation}
\end{equation}
Here, number operator \(\hat n\) is the photon number in the state, \(\theta\) is the random phase angle. The phase distribution is a wrapped normal distribution:

\begin{equation}
f_{WN}(\theta) = \frac{1}{\sigma_\theta \sqrt{2\pi}} \sum_{k=-\infty}^\infty e^{-(\theta + 2\pi k)^2 / 2 \sigma_\theta^2}
\label{eq:wrapped_normal}
\end{equation}
Indeed, the expectation value of the phase shift affects the off-diagonal terms:

\begin{equation}
\rho_{01} \to \langle e^{i \theta} \rangle \rho_{01}, \quad \bar{r} = \int_{-\pi}^\pi f_{WN}(\theta) e^{i \theta} d\theta = e^{-\sigma_\theta^2 / 2}
\label{eq:phase_expectation}
\end{equation}
The \(\bar{r}^2\) factor quantifies the coherence loss, decreasing as \(\sigma_{\theta}\) increases, directly impacting the X-basis measurements in MDI-QKD. Since diagonal terms remain unchanged, phase noise does not contribute to  errors but complicates eavesdropping detection. Consequentely, the combined channel, which integrates thermal loss and phase noise, is:

\begin{equation}
\hat{\rho} \to (1 - \lambda) \begin{bmatrix} \rho_{00} & \bar{r}^2 \rho_{01} \\ \bar{r}^2 \rho_{10} & \rho_{11} \end{bmatrix} + \frac{\lambda}{2} \mathbb{I}
\label{eq:combined_channel}
\end{equation}

\section{ MDI-QKD Analysis: SKR and QBER}
\label{sec3}

In MDI-QKD, Alice and Bob send single-photon states through their respective noisy channels to Charlie, who performs a Bell state measurement, projecting onto: 

\begin{equation}
|\psi^+\rangle = \frac{1}{\sqrt{2}} (|HV\rangle + |VH\rangle), \quad |\psi^-\rangle = \frac{1}{\sqrt{2}} (|HV\rangle - |VH\rangle)
\label{eq:bell_states}
\end{equation}

Assuming single-photon sources and ideal detectors, with Z and X bases each chosen with probability \(\frac{1}{2}\), we analyze the channel effects and derive the secret key rate (SKR) and quantum bit error rate (QBER). We use \( |H\rangle \) (i.e., \( |H\rangle|H\rangle \)) as an illustrative example.

\subsubsection{State Evolution Through Channels}

Alice prepares \( |H\rangle_A \), with density matrix:

\begin{equation}
\hat{\rho}_A = |H\rangle\langle H|_A = \begin{bmatrix} 1 & 0 \\ 0 & 0 \end{bmatrix}
\label{eq:rho_A}
\end{equation}
Bob symmetrically prepares \( |H\rangle_B \), \( \hat{\rho}_B = |H\rangle\langle H|_B \).

\subsubsection{After Channel Transmission}

After passing through the thermal-loss and phase noise channels, the state arrives at Charlie with probability \( P_S^A \) and \( P_S^B \), and the conditional density matrix becomes: 

\begin{equation}
\hat{\rho}_{A'} = (1 - \lambda_A) \hat{\rho}_A + \frac{\lambda_A}{2} \mathbb{I} = \begin{bmatrix} 1 - \frac{\lambda_A}{2} & 0 \\ 0 & \frac{\lambda_A}{2} \end{bmatrix}
\label{eq:rho_A_prime_channel}
\end{equation}
Phase noise does not affect the Z-basis, as it only scales off-diagonal elements. Bob’s state undergoes a similar transformation. The joint density matrix: The joint state is:

\begin{equation}
\hat{\rho}_{A'B'} = \hat{\rho}_{A'} \otimes \hat{\rho}_{B'} 
\label{eq:joint_density}
\end{equation}

\subsubsection{Bell State Measurement}

The projection probability is defined as:

\begin{equation}
P_{\psi^\pm} = \text{Tr} (\hat{\rho}_{A'B'} |\psi^\pm\rangle\langle\psi^\pm|)
\label{eq:projection_prob}
\end{equation}
Accordingly, the probability of \(\hat{\rho}_{A'B'}\) projecting onto \( |\psi^+\rangle \) is:

\begin{align}
P_{\psi^+}^{HH} &= \langle \psi^+ | \hat{\rho}_{A'B'} | \psi^+ \rangle \notag \\
&= \frac{1}{2} \left[ (1 - \frac{\lambda_A}{2}) \frac{\lambda_B}{2} + \frac{\lambda_A}{2} (1 - \frac{\lambda_B}{2}) \right] \notag \\
&= \frac{\lambda_A + \lambda_B - \lambda_A \lambda_B}{4}
\label{eq:proj_psi_plus}
\end{align}
See Appendix A.1 for derivation. 
Similarly, projecting onto \( |\psi^-\rangle \) is:

\begin{equation}
P_{\psi^-}^{HH} = \langle \psi^- | \hat{\rho}_{A'B'} | \psi^- \rangle = \frac{\lambda_A + \lambda_B - \lambda_A \lambda_B}{4}
\label{eq:proj_psi_minus}
\end{equation}
See Appendix A.1 for derivation.
This probability reflects erroneous detection events due to depolarizing noise, impacting the Z-basis error rate. The projection probabilities for various other mixing states can be obtained in the same way.

\subsubsection{QBER and SKR Derivation}

In Table.~\ref{tab:my_table} outlines the coding rules for MDI-QKD. The analysis of error rates and secret key rates is central to evaluating the security of MDI-QKD. The Z-basis is used for key generation, and the error rate reflects the probability of erroneous detection from identical inputs. Accounting for the average gain across all possible inputs, including both valid and invalid detections:
\begin{table}
    \centering
    \caption{MDI-QKD coding rule}
    \label{tab:my_table}
    \begin{tabular}{|c|c|c|} \hline 
         Alice\&Bob & Bell State: \( |\psi^-\rangle \) & Bell State: \( |\psi^+\rangle \) \\ \hline 
         Z-basis & Bit flip & Bit flip \\ \hline 
         X-basis & Bit flip & No Bit flip \\ \hline
    \end{tabular}
\end{table}
\begin{equation}
Q_{\text{Z}} = (1 + \lambda_A \lambda_B) P_S^A P_S^B
\label{eq:Q_Z}
\end{equation}
See Appendix A.2 for explicit derivation.\\
Only different inputs (\( H_A V_B \) and \( V_A H_B \)) contribute to the key, as they produce usable key bits:

\begin{equation}
Q_{\text{Z}}^{1,1} = (2 - \lambda_A - \lambda_B + \lambda_A \lambda_B) P_S^A P_S^B
\label{eq:Q_Z_11}
\end{equation}
We refer the reader to Appendix A.3 for explicit derivation.
Caused by erroneous detection of identical inputs, the Z-basis error rate \(E_Z\) quantifies the probability of projecting onto the Bell state \(|\psi^+\rangle\) when Alice and Bob send identical states such as \(H_A H_B\) or \(V_A V_B\), which should not contribute to the key, and is given by:

\begin{equation}
  E_{\text{Z}} = \frac{\lambda_A + \lambda_B - \lambda_A \lambda_B}{2 (1 + \lambda_A \lambda_B)}
  \label{eq:E_Z}
\end{equation}
See Appendix A.4 for explicit derivation. Where \( E_{\text{Z}} \) quantifies the disturbance due to depolarizing noise in the Z-basis. We note that \(\lambda_A, \lambda_B\) diminishes \(Q_Z\) by elevating the rate of invalid detections, which reduces the SKR and the fraction of valid key bits. In contrast, the X-basis is employed to estimate eavesdropping, influenced by both depolarization and phase noise.

\begin{equation}
  e_{\text{X}}^{1,1}= \frac{1 - (1 - \lambda_A)(1 - \lambda_B) \bar{r}^4}{2}
    \label{eq:e_X_11}
\end{equation}

\begin{strip}
    \begin{align}
R &= P_S^A P_S^B \Bigg[ (2 - \lambda_A - \lambda_B + \lambda_A \lambda_B) \left[ 1 - H\left( \frac{1 - (1 - \lambda_A)(1 - \lambda_B) \bar{r}^4}{2} \right) \right] \\
&\quad - (1 + \lambda_A \lambda_B) f H\left( \frac{\lambda_A + \lambda_B - \lambda_A \lambda_B}{2 (1 + \lambda_A \lambda_B)} \right) \Bigg]
\label{eq:skr}
\end{align}
\end{strip}

See Appendix A.5 and A.6 for derivation. A key figure of merit in quantum security protocols in general, and MDI-QKD in particular, is the SKR. This quantifies the number of secure key bits that Alice and Bob can distill from their correlated measurements, achieved through Charlie’s Bell state measurement, while accounting for potential information leakage to an eavesdropper, Eve. To rigorously formulate this balance, we employ the Devetak-Winter bound \cite{b19} under reverse reconciliation. Based on Eq. \eqref{eq:Q_Z},  Eq. \eqref{eq:Q_Z_11}, Eq. \eqref{eq:E_Z} and Eq. \eqref{eq:e_X_11}, the SKR is expressed in Eq. \eqref{eq:skr}. Here, \(I(A:B) = Q_{Z}^{1,1} [1 - H(e_{X}^{1,1})]\) represents the mutual information between Alice (A) and Bob (B), encapsulating the amount of shared information available for key generation after Charlie’s measurements. The factor \(1 - H(e_{X}^{1,1})\) adjusts the effective key rate by subtracting the information that could potentially be leaked to Eve through errors in the X-basis, as she might exploit these discrepancies to infer key bits, thereby leaving \(I(A:B)\) as the secure information content per valid detection.

The Holevo information, \(\chi(B:E) = Q_Z f H(E_Z)\), quantifies the upper bound on Eve’s knowledge about Bob’s key bits, representing the information she could theoretically extract from the system. We set the parameter \(f = 1\) as ideal error correction efficiency, meaning all Z-basis errors are perfectly reconciled, a simplifying assumption for this analysis. Thus, \(\chi(B:E)\) represents the information leakage Eve could access by observing all Z-basis outcomes, weighted by the entropy of errors that must be corrected to ensure key security.
This Devetak-Winter formulation ensures the SKR accounts for both the usable shared information and Eve’s potential knowledge, effectively mitigating the impacts of thermal noise \(\text{N}_\text{th}\) and phase noise  \(\bar{r}^2\) on the key generation process in MDI-QKD.
This section has established a theoretical model for MDI-QKD that incorporates the effects of thermal-loss and phase-noise channels. By capturing the combined influence of thermal noise (\(N_{\text{th}}\)) and phase noise (through \(\bar{r}^2\)) on the secret key rate, the model lays the groundwork for the simulation-based performance analysis presented in the next section.

\section{Simulation and Analysis of   Channel Noise Effects on MDI-QKD Performance}
\label{sec4}
\begin{figure*}[t!]
\centerline{\includegraphics[width=1\linewidth]{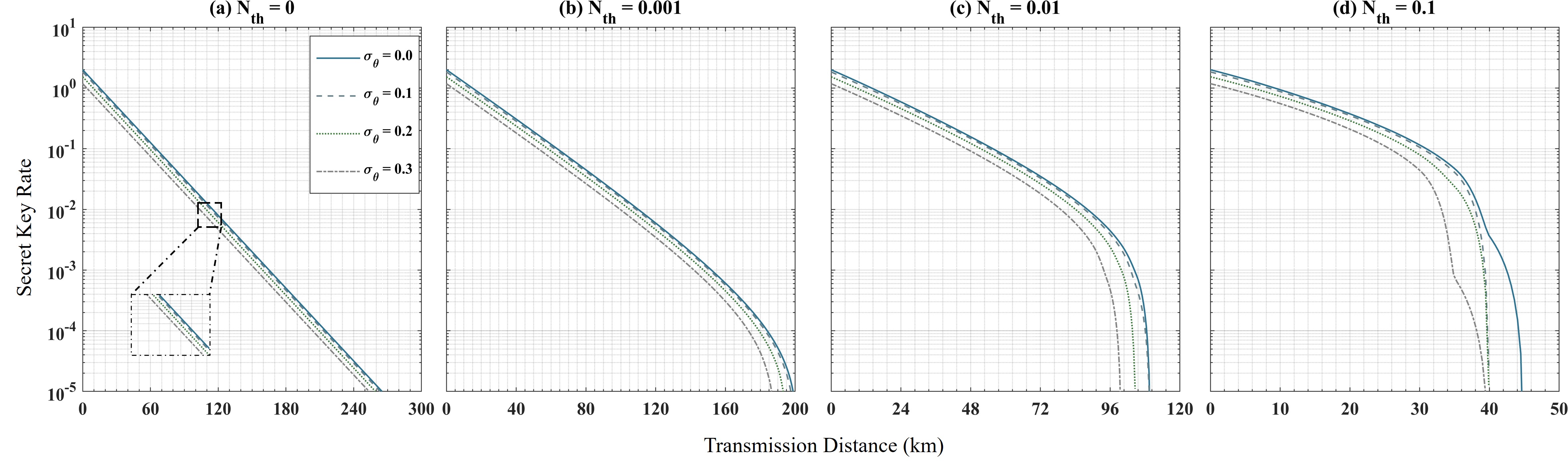}}
\caption{SKR as a function of transmission distance under varying thermal noise (\(N_{\text{th}}\)) and phase noise (\(\sigma_{\theta}\)) in the MDI-QKD system, with \(N_{\text{th}}\) ranging from 0 to 0.1 and \(\sigma_{\theta}\) from 0 to 0.3, highlighting the dominant impact of thermal noise on secure distance reduction.}
\label{fig2}
\end{figure*}
\begin{figure*}[htbp]
\centerline{\includegraphics[width=1\linewidth]{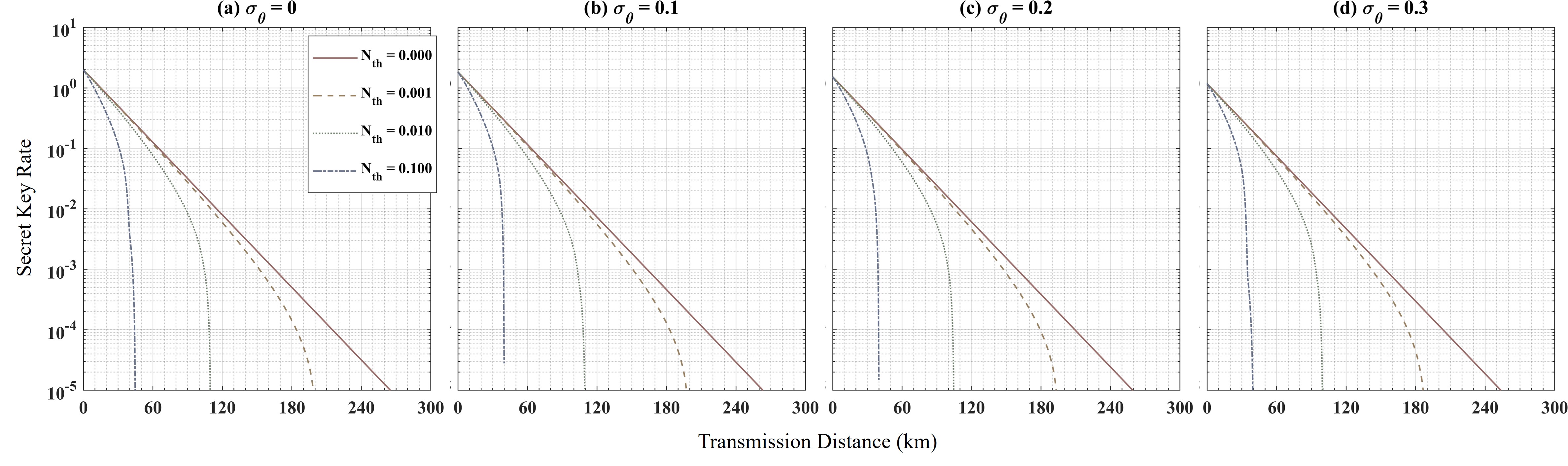}}
\caption{SKR as a function of transmission distance under varying phase noise (\(\sigma_{\theta}\)) and thermal noise (\(N_{\text{th}}\)) in the MDI-QKD system, with \(\sigma_{\theta}\) ranging from 0 to 0.3 and \(N_{\text{th}}\) from 0 to 0.1, showing the limited impact of phase noise compared to thermal noise.}
\label{fig3}
\end{figure*}
In the previous section, we established the MDI-QKD model under thermal-loss and phase-noise channels, forming the basis for the simulation framework. A key parameter in the model is the transmissivity \(\eta\), which quantifies channel efficiency and depends on the physical transmission medium. While the model is general, in our simulations we compute \(\eta\) for a fiber-optic channel using the standard exponential attenuation model:
\(\eta = 10^{-\alpha L / 10}\),
where  \(\alpha=0.2dB/km\),  denotes the fiber attenuation coefficient and \(L\) represents the total transmission distance between Alice, Bob, and Charlie.

Upon completing the simulation setup, we evaluate the SKR across a range of thermal noise levels \(N_{\text{th}}\) and phase noise parameters \(\sigma_{\theta}\), over varying transmission distances. This section presents and analyzes the simulation results to provide deeper insight into how these noise sources influence system performance.

Figure~\ref{fig2} examines the impact of thermal noise \(N_{\text{th}}\) on SKR for several values of phase noise \(\sigma_{\theta}\). The results reveal that SKR is highly sensitive to thermal noise: as \(N_{\text{th}}\) increases, the maximum transmission distance supporting nonzero SKR significantly decreases. Even small increases in thermal noise lead to noticeable reductions in both SKR magnitude and secure distance. In contrast, variations in phase noise within the considered range have only a marginal effect on SKR, especially at longer distances. This suggests that the overall system performance is predominantly constrained by thermal noise, with phase noise playing a secondary role.

Figure~\ref{fig3} reverses the focus, illustrating how SKR responds to increasing phase noise \(\sigma_{\theta}\) across different levels of thermal noise. While SKR shows slight degradation as \(\sigma_{\theta}\) increases, particularly at short to medium distances, the overall sensitivity remains low within the practical phase noise range. In scenarios with negligible thermal noise, increasing \(\sigma_{\theta}\) leads to a modest decline in SKR, reflecting increased phase uncertainty. However, once thermal noise is present at non-negligible levels, it dominates the SKR decay, rendering the system relatively insensitive to further phase noise variations.

The underlying reason for this disparity lies in the respective effects of each noise type on the system’s key rate components. Phase noise introduces a multiplicative attenuation factor \(r^2 = \exp(-\sigma_{\theta}^2)\), which affects the X-basis QBER (\(e_X^{11}\)). However, this factor decreases gradually with \(\sigma_{\theta}\), resulting in only moderate increases in error rates. On the other hand, thermal noise alters the photon-number statistics of the source states, directly impacting the yields \(\lambda_A\) and \(\lambda_B\), and subsequently degrading the gain (\(Q_Z^{11}, Q_Z\)) and increasing the bit error rate \(E_Z\). These combined effects substantially reduce the SKR.

Overall, the simulation results clearly demonstrate that thermal noise is the dominant limiting factor for SKR in MDI-QKD systems, significantly constraining the achievable secure transmission distance. In contrast, phase noise within realistic operating conditions exerts only a limited influence, indicating a degree of robustness to phase fluctuations in optical fibers. These findings highlight the importance of minimizing thermal noise in practical implementations to preserve key rate performance and extend operational range.

\section{Conclusion}
\label{sec5}
In this work, we investigated the performance of MDI-QKD over thermal-loss and phase-noise channels, assuming ideal single-photon sources and perfect detector efficiencies. Employing the Devetak–Winter bound under reverse reconciliation, we derived the secret key rate (SKR) to assess the system’s performance. Our findings indicate that thermal noise has a more pronounced impact on SKR than phase noise, playing a dominant role in limiting the overall efficiency of the MDI-QKD protocol. These analytical insights provide a foundation for future studies aimed at exploring practical MDI-QKD implementations across a broad range of channel conditions.

 \section*{Acknowledgment}
This work was supported by the project Lux4QCI (GA
101091508) funded by the Digital Europe Program, and the
project LUQCIA Funded by the European Union – Next Generation EU, with the collaboration of the Department of Media, Connectivity and Digital Policy of the Luxembourgish Government in the framework of the RRF program. H.P and S.K thank Leonardo Oleynik for discussion. 

\vspace{12pt}

\begin{appendices}

\section{}
\label{app:derivation}

In MDI-QKD, the secure exchange of cryptographic keys between Alice and Bob relies on Charlie’s Bell state measurements. The precision of these measurements, encapsulated in projection probabilities, forms the cornerstone to obtain critical performance metrics such as QBER and SKR,  which collectively quantify the ability of the system. This appendix provides a detailed derivation of the Z-basis projection probabilities \(P_{\psi^+}^{HH}\) and \(P_{\psi^-}^{HH}\) and the X-basis projection probability \(P_{\psi^+}^{DD}\) as representative examples, illustrating the distinct impacts of depolarizing noise (via \(\lambda\)) and phase noise (via \(\bar{r}^2\)). Subsequently, we derive the total count rate \(Q_Z\) , effective key rate \(Q_Z^{1,1}\), Z-basis error rate \(E_Z\) , X-basis error rate \(e_X^{1,1}\) and  \(R\),  using  these projection probabilities. 

\subsection*{A.1 Derivation of Bell State Projection Probabilities}
\label{subsec:proj_prob}

Charlie’s Bell state measurements project the joint state \(\hat{\rho}_{A'B'}\) onto the Bell states \(|\psi^+\rangle\) or \(|\psi^-\rangle \). These probabilities are pivotal, as they reflect how thermal noise (via \(\lambda\)) and phase noise (via \(\bar{r}^2\)) alter the quantum state, influencing key generation and error rates. As examples, we derive \(P_{\psi^+}^{HH}\) and \(P_{\psi^-}^{HH}\) for the identical input of the Z-basis \(|H\rangle_A |H\rangle_B\), and \(P_{\psi^+}^{DD}\) for the identical input of the X-basis \(|D\rangle_A |D\rangle_B\), while summarizing the remaining probabilities in A.1.3 to provide a complete overview.

\subsubsection*{A.1.1 Z-basis: \(P_{\psi^+}^{HH}\) and  \(P_{\psi^-}^{HH}\)}

For the input \(|H\rangle_A |H\rangle_B\), Alice and Bob prepare horizontally polarized photons, forming the initial state \(\hat{\rho}_A \otimes \hat{\rho}_B\). After transmission through independent noisy channels the indivudual states in Eq. \eqref{eq:combined_channel}, this evolves into outcoming states states \(\hat{\rho}_{A'}\) and \(\hat{\rho}_{B'}\). According to Eq. \eqref{eq:joint_density}, the joint density matrix \(\hat{\rho}_{A'B'} = \hat{\rho}_{A'} \otimes \hat{\rho}_{B'}\) in the Z-basis is:

\begin{equation}
\begin{aligned}
&\hat{\rho}_{A'B'} = \hat{\rho}_{A'} \otimes \hat{\rho}_{B'} \\
&= {\scriptsize \begin{bmatrix}
\left(1 - \frac{\lambda_A}{2}\right)\left(1 - \frac{\lambda_B}{2}\right) & 0 & 0 & 0 \\
0 & \left(1 - \frac{\lambda_A}{2}\right) \frac{\lambda_B}{2} & 0 & 0 \\
0 & 0 & \frac{\lambda_A}{2} \left(1 - \frac{\lambda_B}{2}\right) & 0 \\
0 & 0 & 0 & \frac{\lambda_A}{2} \frac{\lambda_B}{2}
\end{bmatrix}}
\end{aligned}
\label{eq:rho_Aprime_Bprime_matrix}
\end{equation}
This diagonal form reflects the Z-basis input’s lack of superposition, rendering phase noise (\(\bar{r}^2\)) ineffective here, as it only affects off-diagonal terms. Depolarizing noise (\(\lambda_A, \lambda_B\)), however, mixes polarizations, potentially causing errors in detection.
The Bell state \(|\psi^+\rangle\) and \(|\psi^-\rangle\) are expressed in vector form:
\begin{equation}
\begin{aligned}
|\psi^+\rangle &= \frac{1}{\sqrt{2}} \begin{bmatrix} 0 \\ 1 \\ 1 \\ 0 \end{bmatrix}, &
\langle \psi^+ | &= \frac{1}{\sqrt{2}} \begin{bmatrix} 0 & 1 & 1 & 0 \end{bmatrix}
\end{aligned}
\label{eq:psi_plus}
\end{equation}
\begin{equation}
\begin{aligned}
|\psi^-\rangle &= \frac{1}{\sqrt{2}} \begin{bmatrix} 0 \\ 1 \\ -1 \\ 0 \end{bmatrix}, &
\langle \psi^- | &= \frac{1}{\sqrt{2}} \begin{bmatrix} 0 & 1 & -1 & 0 \end{bmatrix}
\end{aligned}
\label{eq:psi_minus}
\end{equation}
The projection probability onto \(|\psi^+\rangle\),  as given by Eq.\eqref{eq:proj_psi_plus}, can be formulated using the vector representation of Eq.\eqref{eq:psi_plus} as follows:
    \[
    P_{\psi^+}^{HH} = \frac{1}{2} [0 \, 1 \, 1 \, 0] \hat{\rho}_{A'B'} \begin{bmatrix} 0 \\ 1 \\ 1 \\ 0 \end{bmatrix}
    \]
Matrix operations with the given  Eq. \eqref{eq:rho_Aprime_Bprime_matrix} produce the result given in  Eq. \eqref{eq:proj_psi_plus} as:
    \[
    P_{\psi^+}^{HH} = \frac{\lambda_A + \lambda_B - \lambda_A \lambda_B}{4}
    \]
This non-zero probability signifies an erroneous detection event where identical inputs  \(|H\rangle_A |H\rangle_B\) are mistaken for \(|\psi^+\rangle\), a state typically associated with different inputs. This error, induced by depolarizing noise, increases with \(\lambda_A\) and \(\lambda_B\), directly contributing to the error rate on the Z-basis and necessitating error correction, which reduces the yield of the secure key.
Similarly, the projection probability onto \(|\psi^-\rangle\) in Eq. \eqref{eq:proj_psi_minus} is expressed as:
\[
P_{\psi^-}^{HH} = \frac{1}{2} [0 \, 1 \, -1 \, 0] \hat{\rho}_{A'B'} \begin{bmatrix} 0 \\ 1 \\ -1 \\ 0 \end{bmatrix}
\]
Matrix operations with the given  Eq. \eqref{eq:rho_Aprime_Bprime_matrix} produce the result given in Eq. 
This probability, identical to \(P_{\psi^+}^{HH}\), indicates that identical inputs contribute equally to both Bell states, amplifying the Z-basis error rate due to thermal noise (\(\lambda_A, \lambda_B\)).
\eqref{eq:proj_psi_minus} as:
\[
P_{\psi^-}^{HH} = \frac{1}{2} \left[ \left(1 - \frac{\lambda_A}{2}\right) \frac{\lambda_B}{2} - \frac{\lambda_A}{2} \left(1 - \frac{\lambda_B}{2}\right) \right] = \frac{\lambda_A + \lambda_B - \lambda_A \lambda_B}{4}
\]

\subsubsection*{A.1.2 X-Basis: \(P_{\psi^+}^{DD}\)}
For the X-basis input \(|D\rangle_A |D\rangle_B\), the joint density matrix after transmission through both channels, as described in Eq. \eqref{eq:combined_channel}, is expressed as:
\begin{strip}
\begin{equation}
\begin{aligned}
&\hat{\rho}_{A'B'} = \hat{\rho}_{A'} \otimes \hat{\rho}_{B'} \\
&= \begin{bmatrix} 
    \frac{1}{4} & \frac{(1 - \lambda_B)\bar{r}^2}{4} & \frac{(1 - \lambda_A)\bar{r}^2}{4} & \frac{(1 - \lambda_A)(1 - \lambda_B)\bar{r}^4}{4} \\
    \frac{(1 - \lambda_B)\bar{r}^2}{4} & \frac{1}{4} & \frac{(1 - \lambda_A)(1 - \lambda_B)\bar{r}^4}{4} & \frac{(1 - \lambda_A)\bar{r}^2}{4} \\
    \frac{(1 - \lambda_A)\bar{r}^2}{4} & \frac{(1 - \lambda_A)(1 - \lambda_B)\bar{r}^4}{4} & \frac{1}{4} & \frac{(1 - \lambda_B)\bar{r}^2}{4} \\
    \frac{(1 - \lambda_A)(1 - \lambda_B)\bar{r}^4}{4} & \frac{(1 - \lambda_A)\bar{r}^2}{4} & \frac{(1 - \lambda_B)\bar{r}^2}{4} & \frac{1}{4}
\end{bmatrix}
\end{aligned}
\label{Annexeq:joint_state_DD}
\end{equation}
\end{strip}

This matrix, derived from the tensor product of individual states \(\hat{\rho}_{A'}\) and \(\hat{\rho}_{B'}\), exhibits off-diagonal terms scaled by \(\bar{r}^2\), reflecting phase noise’s impact on the initial superposition, alongside depolarizing noise’s mixing effects parameterized by \(\lambda_A\) and \(\lambda_B\).
Similarly, performing matrix operations, projecting onto \(|\psi^+\rangle\) yields:
\begin{equation}
P_{\psi^+}^{DD} = \frac{1 + (1 - \lambda_A)(1 - \lambda_B) \bar{r}^4}{4}
\label{eq:P_psi_plus_DD}
\end{equation}
We note that Phase noise (\(\bar{r}^2 < 1\)) reduces the off-diagonal contribution, lowering \(P_{\psi^+}^{DD}\) and increasing X-basis errors, critical for eavesdropping detection.

\subsubsection*{A.1.3 Summary of all projection probabilities}
\label{subsubsec:proj_prob_all}
With similar derivation methods, the remaining probabilities are provided for direct use:
   \begin{align*}
P_{\psi^+}^{HH} = P_{\psi^-}^{HH} &= P_{\psi^+}^{VV} = P_{\psi^-}^{VV} = \frac{\lambda_A + \lambda_B - \lambda_A \lambda_B}{4} \\
P_{\psi^+}^{HV} = P_{\psi^+}^{VH} &= \frac{2 - \lambda_A - \lambda_B + \lambda_A \lambda_B}{4} \\
P_{\psi^-}^{HV} = P_{\psi^-}^{VH} &= \frac{2 - \lambda_A - \lambda_B + \lambda_A \lambda_B}{4} \\
P_{\psi^+}^{AA} = P_{\psi^+}^{DD} &= \frac{1 + (1 - \lambda_A)(1 - \lambda_B) \bar{r}^4}{4} \\
P_{\psi^-}^{AA} = P_{\psi^-}^{DD} &= \frac{1 - (1 - \lambda_A)(1 - \lambda_B) \bar{r}^4}{4} \\
P_{\psi^+}^{AD} = P_{\psi^+}^{DA} &= \frac{1 - (1 - \lambda_A)(1 - \lambda_B) \bar{r}^4}{4} \\
P_{\psi^-}^{AD} = P_{\psi^-}^{DA} &= \frac{1 + (1 - \lambda_A)(1 - \lambda_B) \bar{r}^4}{4}
\end{align*}

\subsection*{A.2 Derivation of Z-basis total count rate  \(Q_Z\) }
\label{subsec:Q_Z}

The total count rate \(Q_Z\) quantifies the average probability that Charlie detects a Bell state on the Z-basis across all possible input combinations (\(H_AH_B, V_AV_B, H_AV_B, V_AH_B\)). This metric encompasses both valid key-generating events (e.g., \(H_A V_B\) and \(V_A H_B\)) and invalid detections (e.g., \(H_AH_B\) and \(V_AV_B\)) that contribute to errors. In MDI-QKD, Charlie’s measurement projections onto \(|\psi^+\rangle\) or \(|\psi^-\rangle\) as valid responses, regardless of whether they correspond to the intended key bits, reflecting the overall detection performance of the system. Influenced primarily by depolarizing noise (\(\lambda_A, \lambda_B\)), \(Q_Z\) is crucial for assessing the measurement efficiency and noise impact:
\begin{equation}
\begin{aligned}
    Q_Z &= \frac{1}{2} P_S^A P_S^B\\
   &\left( P_{\psi^+}^{HH} + P_{\psi^+}^{VV} + P_{\psi^+}^{HV} + P_{\psi^-}^{HV} + P_{\psi^+}^{VH} + P_{\psi^-}^{VH} \right)
\end{aligned}
\label{Annexeq:Q_Z_intermediate}
\end{equation}
Here, the factor \(\frac{1}{2}\) averages over the equal probabilities of Alice and Bob selecting \(|H\rangle\) or \(|V\rangle\), and \(P_S^A P_S^B\) represents the joint success probability of photon transmission through the noisy channels. By including projections in both Bell states, \(|\psi^+\rangle\) and \(|\psi^-\rangle\), \(Q_Z\) accounts for all detection events, whether valid key-generating outcomes or erroneous detections, shaping the system performance under depolarizing noise.
Substituting the probabilities listed in A.1.3 into Eq. \eqref{Annexeq:Q_Z_intermediate}, the result reflects the interplay between successful photon transmissions (\(P_S^A P_S^B\)) and noise-induced detections, which produces Eq. \eqref{eq:Q_Z} :
\[
Q_Z = (1 + \lambda_A \lambda_B) P_S^A P_S^B
\]
An increase in \(Q_Z\) with rising \(\lambda_A\) and \(\lambda_B\) signifies a greater proportion of invalid detections, requiring robust error correction to maintain security.

\subsection*{A.3 Derivation of Z-basis effective key rate \(Q_Z^{1,1}\)}
\label{subsec:Q_Z_11}

The effective key rate \(Q_Z^{1,1}\) isolates the probability of detecting Bell states from different inputs (\(H_A V_B, V_A H_B\)), which generate usable key bits critical for the secure keys in MDI-QKD. Unlike the total count rate, this metric focuses solely on valid events that align with the protocol’s coding rules in Table.~\ref{tab:my_table}, directly contributing to the mutual information \(I(A:B)\) in the SKR. It is expressed as:
\begin{equation}
Q_Z^{1,1} = \frac{1}{2} \left( P_{\psi^+}^{HV} + P_{\psi^-}^{HV} + P_{\psi^+}^{VH} + P_{\psi^-}^{VH} \right) P_S^A P_S^B
\label{Annexeq:Q_Z_11_intermediate}
\end{equation}
Here, the factor \(\frac{1}{2}\) averages over the equal probabilities of Alice and Bob selecting opposite polarization states (\(|H\rangle_A |V\rangle_B\) or \(|V\rangle_A |H\rangle_B\)).
Substituting the probabilities from A.1.3 into Eq. \eqref{Annexeq:Q_Z_11_intermediate}, the result captures only valid detection events, sensitive to the effect of depolarizing noise on reducing the state distinguishability, which produces Eq. \eqref{eq:Q_Z_11} :
\[
Q_Z^{1,1} = (2 - \lambda_A - \lambda_B + \lambda_A \lambda_B) P_S^A P_S^B
\]
A decrease in \(Q_Z^{1,1}\) with increasing \(\lambda_A\) and \(\lambda_B\) highlights the adverse impact of thermal noise on key generation efficiency.

\subsection*{A.4 Derivation of Z-basis error rate \(E_Z\)}
\label{subsec:E_Z}

The Z-basis error rate \(E_Z\) quantifies the fraction of erroneous detections where identical inputs (\(H_AH_B, V_AV_B\)) are mistaken for valid Bell states, a consequence of the randomization of the polarization states by thermal noise. This metric is pivotal for determining error correction costs, as it measures the noise-induced deviations that must be reconciled to ensure key consistency:
\begin{equation}
E_Z = \frac{P_{\psi^+}^{HH} + P_{\psi^+}^{VV}}{P_{\psi^+}^{HH} + P_{\psi^+}^{VV} + P_{\psi^+}^{HV} + P_{\psi^-}^{HV} + P_{\psi^+}^{VH} + P_{\psi^-}^{VH}}
\label{Annexeq:E_Z_intermediate}
\end{equation}
Here, the numerator sums the probabilities of incorrect projections onto \(|\psi^+\rangle\) from identical inputs, while the denominator, equivalent to \(Q_Z\) Eq. \eqref{eq:Q_Z}, represents the total detection probability in all input combinations of the Z-basis.
Substituting the probabilities from A.1.3 into Eq. \eqref{Annexeq:E_Z_intermediate}, the result reflects the impact of depolarizing noise, which yields Eq. \eqref{eq:E_Z} from the main text:
\[
E_Z = \frac{\lambda_A + \lambda_B - \lambda_A \lambda_B}{2 (1 + \lambda_A \lambda_B)}
\]
The dependence on \(\lambda_A\) and \(\lambda_B\) underscores the role of thermal noise in increasing error rates, thus reducing the yield of secure keys by increasing the resources needed for error correction.

\subsection*{A.5 Derivation of X-basis error rate \(e_X^{1,1}\) }
\label{subsec:e_X_11}
The X-basis error rate \(e_X^{1,1}\) assesses the potential for eavesdropping by measuring the frequency of incorrect detections in the X-basis, influenced by both depolarizing noise (\(\lambda_A, \lambda_B\)) and phase noise (\(\bar{r}^2\)). This metric is essential for security analysis, as it indicates deviations from expected coherent outcomes that an eavesdropper might exploit:
\begin{equation}
e_X^{1,1} = \frac{P_{\psi^-}^{DD} + P_{\psi^+}^{DA}}{P_{\psi^+}^{DD} + P_{\psi^-}^{DD} + P_{\psi^+}^{DA} + P_{\psi^-}^{DA}}
\label{Annexeq:e_X_11_intermediate}
\end{equation}
Align with the protocol’s coding rules in Table.~\ref{tab:my_table}, the numerator captures error events where \(|D\rangle_A |D\rangle_B\) projects onto \(|\psi^-\rangle\) or \(|D\rangle_A |A\rangle_B\) onto \(|\psi^+\rangle\), while the denominator sums all possible X-basis detection probabilities.

Using the probabilities from A.1.3 in Eq. \eqref{Annexeq:e_X_11_intermediate}, resulting in Eq. \eqref{eq:e_X_11} :
\[
e_X^{1,1} = \frac{1 - (1 - \lambda_A)(1 - \lambda_B) \bar{r}^4}{2}
\]

An increase in \(e_X^{1,1}\) with a decrease in \(\bar{r}^2\) indicates the critical role of phase noise in security analysis.

\subsection*{A.6 Derivation of SKR \(R\)}
\label{subsec:skr}
The SKR: \(R\) quantifies secure key bits, balancing key generation with leakage via the Devetak-Winter bound in:
\begin{equation}
R = I(A:B) - \chi(B:E) = Q_{Z}^{1,1} [1 - H(e_{X}^{1,1})] - Q_Z f H(E_Z)
\label{Annexeq:R_intermediate}
\end{equation}
Here, \(Q_Z^{1,1} [1 - H(e_X^{1,1})]\) represents the secure key fraction after accounting for potential eavesdropping losses in the X basis, while \(Q_Z H(E_Z)\) quantifies the information lost to error correction in the Z-basis. The binary entropy function \(H(x) = -x \log_2(x) - (1 - x) \log_2(1 - x)\) (for \(0 < x < 1\), with \(H(0) = H(1) = 0\)) measures the uncertainty of the error probabilities \(e_X^{1,1}\) and \(E_Z\), reflecting the amount of information Eve could gain or must be corrected for.

Combining the results of A.2, A.3, A.4 and A.5 into Eq. \eqref{Annexeq:R_intermediate}, the final expression reflects the interplay of noise effects, which produces Eq. \eqref{eq:skr}:
\begin{strip}
\begin{equation}
\begin{aligned}
R &= P_S^A P_S^B \Bigg[ (2 - \lambda_A - \lambda_B + \lambda_A \lambda_B) \left[ 1 - H\left( \frac{1 - (1 - \lambda_A)(1 - \lambda_B) \bar{r}^4}{2} \right) \right] \\
&\quad - (1 + \lambda_A \lambda_B) f H\left( \frac{\lambda_A + \lambda_B - \lambda_A \lambda_B}{2 (1 + \lambda_A \lambda_B)} \right) \Bigg]
\end{aligned}
\end{equation}
\end{strip}

We highlight that , thermal noise dominates SKR reduction via \(Q_Z\) and \(E_Z\), while phase noise affects \(e_X^{1,1}\), shaping security trade-offs, as is detailed in the text.

\end{appendices}
\end{document}